\documentclass[prl,showpacs,showkeys,amsmath,amssymb,superscriptaddress,twocolumn] {revtex4}
\usepackage{graphicx}
\usepackage{amsfonts}

\def\rc{\phi_{\rm rcp}}
\def\rl{\phi_{\rm rlp}}
\def\vrl{\phi_{\rm rvlp}}

\begin{document}
\title{Disordered jammed packings of frictionless spheres}

\author{Massimo Pica Ciamarra}
\affiliation{
CNR--SPIN, Dipartimento di Scienze Fisiche,
Universit\'a di Napoli Federico II, Italy
}

\author{Antonio Coniglio}
\affiliation{
CNR--SPIN, Dipartimento di Scienze Fisiche,
Universit\'a di Napoli Federico II, Italy
}

\author{Antonio de Candia}
\affiliation{
CNR--SPIN, Dipartimento di Scienze Fisiche,
Universit\'a di Napoli Federico II, Italy
}

\begin{abstract}
At low volume fraction, disordered arrangements of frictionless spheres are found in un--jammed states unable to support applied stresses, while at high volume fraction they are found in jammed states with mechanical strength. Here we show, focusing on the hard sphere zero pressure limit, that the transition between un-jammed and jammed states does not occur at a single value of the volume fraction, but in a whole volume fraction range.
This result is obtained via the direct numerical construction of disordered jammed states with a volume fraction varying between two limits, $0.636$ and $0.646$. We identify these limits with the random loose packing volume fraction $\rl$ and the random close packing volume fraction $\rc$ of frictionless spheres, respectively.
\end{abstract}
\maketitle

\section{Introduction}
Packing problems are among the most ancient. 
For instance, about 2200 years ago Archimedes faced the problem of counting the number of grains of the
beaches of its home town, Syracuse, and succeeded in demonstrating that this number is finite (Archimedes, {\it The sand rockener}). Another eminent Greek scientists, Apollonious of Perga, is also renowned for his works on disk and sphere packings.
Modern works on disordered packing of grains dates back to the experiments conducted by
Bernal\cite{Bernal1960} and Scott\cite{Scott1960}, who prepared
packings of monodisperse spheres using different protocols, and were able to measure their volume fraction $\phi$, defined as the fraction of the total volume occupied by the spheres. 
Their results indicated the existence of an upper and of a lower bound for the volume fraction of stable disordered arrangements of
spheres,  named `Random close packing', $\rc$, and `Loose random packing', $\rl$. Bernal estimated $\rc = 0.63 \pm 0.07$ and $\rl = 0.60 \pm 0.01$, in  the infinite system size limit. 
Regarding the existence of any first principle definition of these two bounds, Bernal\cite{Bernal1960} speculated that  `The  figure for the occupied volume of random close packing must be mathematically determinable, although so far we known
undetermined'. Conversely, he questioned the existence of a first principle definition of the lower bound $\rl$: `The
mathematical status of physical random loose packing is not so evident'.

Subsequent works clarified that $\rl$ depends on the Coulomb friction coefficient. Onoda and Liniger\cite{Onoda1990} operatively defined
$\rl$ as the smallest volume fraction attainable letting the particle sediment under gravity. This lower bound is obtained in the limit of
zero sedimentation velocity, when particles fall in a very high viscous fluid. Onoda and Liniger\cite{Onoda1990} estimated $\rl = 0.555 \pm
0.005$, while a more recent work\cite{RLP_matthias} suggests $\rl = 0.550 \pm 0.001$.

Since $\rl$ depends on the friction coefficient, it is of interest to consider its value in the frictionless case.
Published results seem to indicate that, in absence of friction, $\rl = \rc$.
Consider, for instance, the `jamming phase diagram' introduced by Liu and Nagel\cite{Liu1998}, illustrating the region of the temperature, volume fraction and stress space where jammed (mechanically stable) systems are found. 
Along the volume fraction axis, i.e. at zero temperature and zero applied stress, the transition between the unjammed and the jammed phase is marked  to occur at a single value of the volume fraction, known as the J--point and later identified with the random close packing volume fraction  $\rc$. Since for $\phi < \rc$ there are not jammed states, the diagram suggests that at zero friction $\rl$ coincides with $\rc$. 
The same conclusion could be drawn from the results of O'Hern and coworkers\cite{OHern2002,OHern2003}, who numerically generated jammed packings of soft frictionless spheres using the conjugate--gradient protocol. 
In the hard sphere limit, they found jammed packings only at the volume fraction $\phi^* = 0.639 \pm 0.01$, identified with $\rc$.
Being the only volume fraction at which jammed packings are found, it seems obvious to also identify $\phi^*$ with $\rl$, concluding that at zero friction $\rc = \rl$.

In this manuscript we give evidence that, at zero friction, $\rl < \rc$. An indication suggesting this
possibility comes from the comparison of numerical results found by different research groups, which have used slightly different algorithms to prepare jammed packings of frictionless particles. For instance, using the conjugate gradient method\cite{OHern2002,OHern2003}, O'Hern et al. suggested $\rc \simeq
0.639$, while using a packing inflation algorithm Zhang and Makse\cite{Makse2005} obtained $\rc \simeq 0.645$ (with an error of the order of $10^{-5}$). A very close value, $0.644$, was also reported in previous works\cite{Scott1969,Berryman1983,Jodrey1985}. These estimates are
close, but not consistent within the reported errors.
The discrepancies could be in principle attributed to one of the following causes:
\begin{enumerate}
\item[A:] Finite--size effects. One or both estimates are wrong as affected by finite--size effects.
\item[B:] Ordering. The upper bound for the volume fraction of grain packings is that of the FCC crystal, $\phi_{\rm FCC} \simeq 0.74$. One could therefore speculate that jammed packings with volume fraction above $\rc \simeq 0.639$ contains some crystalline patches.
\item[C:] $\rl < \rc$. Disordered jammed packings occur in a whole volume fraction range, at least varying from $0.639$ to $0.645$.
\end{enumerate}
The possibility A) must be excluded, as finite--size are known to influence the jamming volume fraction leading to a smaller estimate of the critical packing fraction\cite{Desmond}. However, the smaller of the above estimates, $\rc \simeq 0.639$, has been obtained via a careful study of the infinite system size limit\cite{OHern2002,OHern2003}.
Here we introduce an algorithm able to generate jammed packings in a large volume fraction range (enclosing the range $0.639$--$0.644$), and show that point B must also be excluded. In fact, altough the concept of random close packing of spheres is ill--defined\cite{Torquato2000} due to the absence of precise definition of `randomness', nevertheless our results 
indicate that it is possible to generate jammed disordered state with no cristalline patches up to the volume fraction $\phi \simeq 0.646$, which is our estimation of the random close packing volume fraction.
Accordingly, we suggest that point C above is correct, i.e. that the jamming transition of frictionless spheres occurs along a whole volume fraction range.

The manuscript is organized as follows. In Sec.~\ref{sec:protocol} we describe our numerical model and the protocol used to generated jammed packings of frictionless spheres.
In Sec.\ref{sec:rlp}, following Onoda and Liniger\cite{Onoda1990}, we define as random loose volume fraction $\rl$ the volume fraction which is attained in the limit of infinitesimally slow energy minimization, and determine its value for frictionless spheres in no gravity.
We show that it is possible to generate un--jammed packings with a volume fraction which is higher than $\rl$ in Sec.~\ref{sec:rcp}, and describe their degree of order in Sec.~\ref{sec:order}. This analysis clarifies that disordered un-jammed states cannot be obtained when the volume fractions overcomes a threshold, we identify with the random close packing volume fraction, $\rc > \rl$.
Open questions and future directions are presented in Sec.~\ref{sec:conclusions}.

\section{Protocols\label{sec:protocol}}
In this section, we introduce the numerical model, give the definition of zero pressure jammed packings of spheres, and
describe the protocols used to generate jammed packings of spheres.
\subsection{Numerical model and zero pressure jammed packing of spheres}
We consider a system of monodisperse frictionless soft spheres of diameter $D$, and mass $m$, interacting with an harmonic potential
$v(r_{ij})$,
\begin{equation}
v(r_{ij}) =\left\{ 
\begin{array}{ll}
\epsilon (1-r_{ij}/D)^2 & {\rm if~~} r \leq D, \\
0, & {\rm otherwise},
\end{array}
\right.
\end{equation}
where ${\bf r}_i$ specifies the position of particle $i$, $r_{ij} = \left| {\bf r}_i - {\bf r}_j \right|$ is the distance between particles
$i$ and $j$. $D$, $m$ and $\epsilon$ are our units of length, mass and energy, respectively.
The elastic $E$ energy of this system is
\begin{equation}
E(\{r,\dot{r}\}) = \sum_{i \neq j}^N v(r_{ij}),
\end{equation}
where ${\bf \dot{r}_i}$ is the velocity of particle $i$,  and $N$ the number of particles.

Jammed configurations correspond to energy minima of the system. Here we consider a configuration jammed when the mean energy per particle is higher than a small threshold, $10^{-10}\epsilon$.
In principle, the definition of jamming we are considering allows for the presence of particles with no contacts (rattlers). However, in the infinite system size limit their concentration is known to vanish\cite{OHern2003}.

\begin{figure}[t!]
\begin{center}
\includegraphics*[scale=0.30]{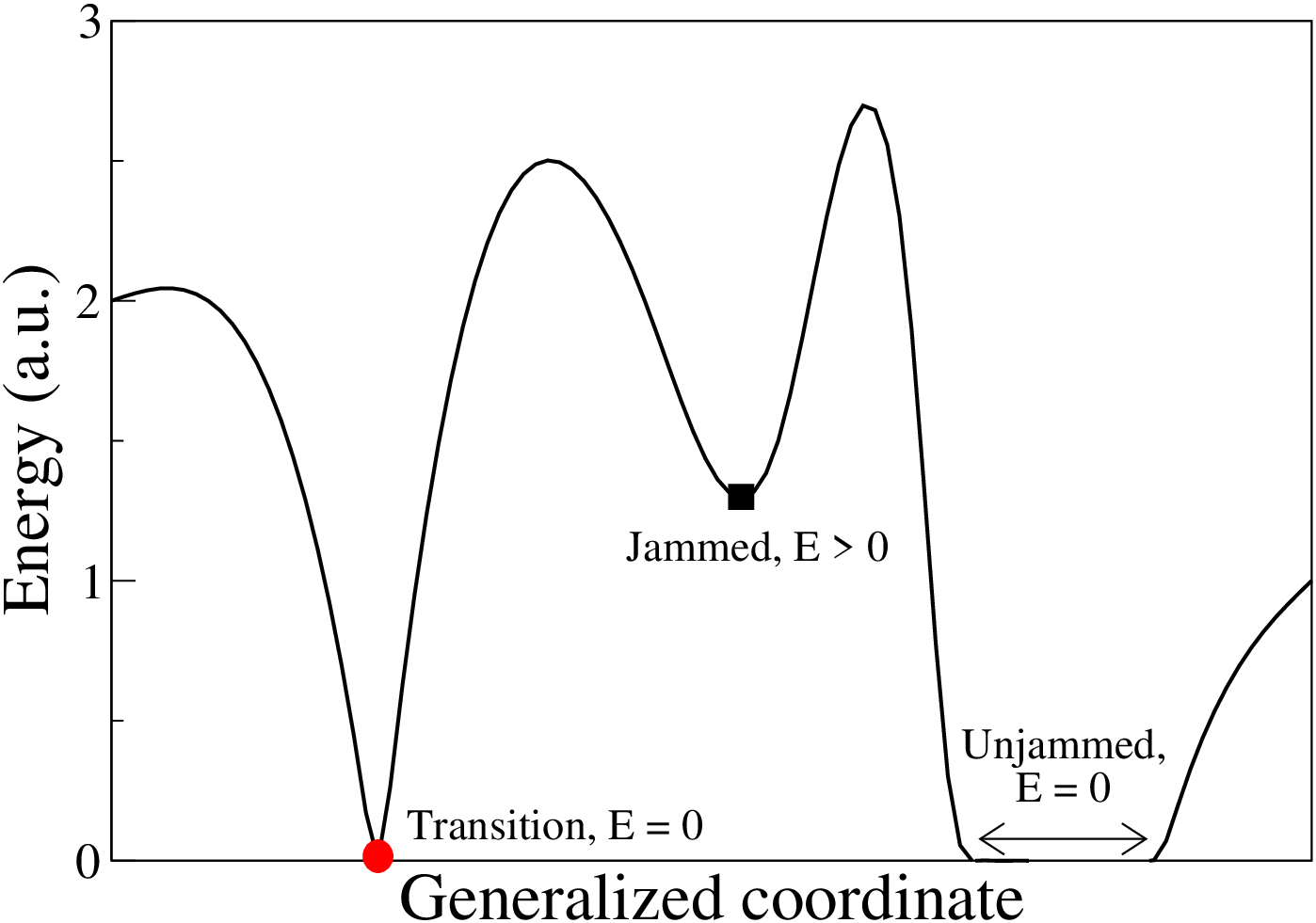}
\end{center}
\caption{\label{fig:minimization} Illustration of the possible states resulting from an energy minimization protocol. These
states can be jammed with a finite energy, jammed with zero energy (transition states), or unjammed with zero energy.
}
\end{figure}

\subsection{Protocols}
Jammed packings of spheres correspond to minima of the energy of the system
$E = \sum_{i \neq j} v(r_{ij})$. We sample these minima solving the equation of motion of
the system in the presence of a viscous damping term, as if the particles were
immersed in a viscous fluid, starting from a random (infinite temperature) configuration.
Each particle evolves according to:
\begin{equation}
\label{eq:dissipation}
m \ddot{r_i} = -\frac{\partial E}{\partial r_i} - \eta \dot{r_i},
\end{equation}
where the parameter $\eta$ plays the role of a viscosity. As time goes on, the total energy of the system decreases because of the
dissipative term, and the dynamics eventually stops (the kinetic energy vanishes). 
As illustrated in Fig.~\ref{fig:minimization}, the resulting state will be either jammed with zero energy, or jammed with a finite energy.
The figure also illustrates the existence of transition states, jammed states of zero energy.
Energy minima may be found using other protocols, as for instance simulated annealing, in which the system is coupled to a heat bath whose temperature is decreased until it vanishes. Alternatively, one could consider the minimization of the energy of the system using as initial state an equilibrium configuration at a given value of the temperature: this is the protocol used to unveil the features of the energy landscape sampled by glass--forming liquids\cite{Stillinger1984,Stillinger1985}.

Qualitatively, the role of the parameter $\eta$ in the energy minimization procedure is easily understood. 
When $\eta$ is high, the kinetic energy of the system is small, which implies that the system is less able to escape from the energy
basins it visits. 
Accordingly, the larger $\eta$ the smaller the region of the configurational space the system explores before getting
trapped in a energy basin. In particular, in the limit $\eta \to \infty$ the system jams in the first minimum it enters. 
Conversely, at a finite value of $\eta$, the system may reach a minimum of the potential energy with a finite value of the kinetic energy, and may therefore be able to escape from it overcoming the confining energy barriers. 
The average value of the energy of the minima reached with a given value of $\eta$ is therefore expected to decrease as $\eta$ decreases.

\section{Random loose packing\label{sec:rlp}}
Onoda and Liniger\cite{Onoda1990} operatively defined $\rl$ as the smallest volume fraction attained via sedimentation
protocols under gravity. Since sedimentation is an energy (gravitational energy) minimization protocol, this operative definition can be extended to the absence of gravity, simply defining $\rl$ as the smallest volume fraction of jammed states obtained via energy minimization protocols, the initial state being a random one.

This jammed state of large volume fraction is obtained solving the equations of motion Eq.~\ref{eq:dissipation} in the limit $\eta \to \infty$, in the infinite time limit.
Operatively, this state can be also obtained via more tractable numerical procedures able to find energy minima, namely the steepest descend method, or the equivalent but computationally more performant conjugate gradient (CG) method. 
Here we use the Fletcher--Reeves CG algorithm, as implemented in the GNU scientific library~\cite{gnu}.
The CG is characterized by two parameters.
First, there is a tolerance small parameter, which is used to decide when the minimization along a given direction of the configurational
space stops. Afterwards, the algorithm tries to minimize the energy moving the system along a conjugate
direction of the phase space.
In the minimization of the energy $E = \sum_{i \neq j}  v(r_{ij})$ we have not found a dependence of the results on this
parameter, when this is small enough.
Second, there is a threshold parameter $\delta e$: the algorithm stops when the energy difference between two successive iterations is smaller than this threshold. Here we set $\delta e = 0$, meaning that the algorithm stops when the energy variation in successive iterations is smaller than our numerical precision. 
We note that zero--energy jammed packings obtained via the CG method have been previously\cite{OHern2002,OHern2003} identified with $\rc$, while here we identify them with $\rl$. We comment on this point later on.

\begin{figure}[t!]
\begin{center}
\includegraphics*[scale=0.30]{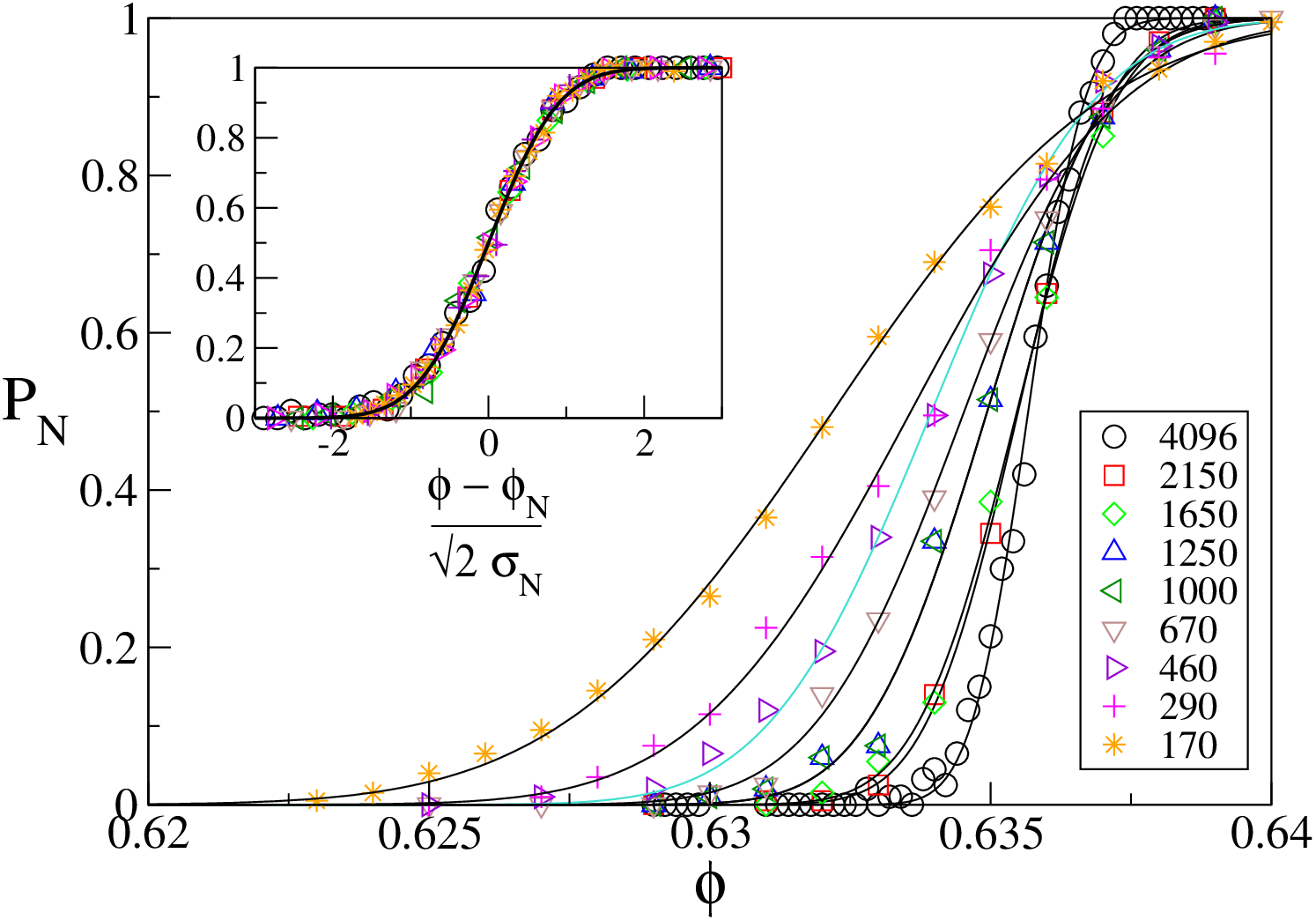}
\end{center}
\caption{\label{fig:fiterf} Probability $P_N$ of obtaining a jammed configuration minimizing the energy via the CG protocol, as a
function of the volume fraction $\phi$, for different values of $N$. Each point is estimated performing $200$ independent CG
minimizations. For each value of $N$, the data are fitted by a scaled error function with inflection point $\phi_N$ and standard deviation
$\sigma_N$ (plain  lines). The inset shows the corresponding data collapse.}
\end{figure}

To determine $\rl$ in the infinite system size limit we have performed a finite-size scaling, considering systems with a number of
particles $N$ varying from $170$ to $4096$. 
For each value of $N$, we considered different values of the volume fraction $\phi$, performed $200$ independent energy minimization CJ protocols, and measured the fraction $P_N(\phi)$ of these minimization procedures yielding jammed configurations. To this end, we have considered a configuration as jammed when its mean energy per particles is greater that $10^{-10}\varepsilon$, unjammed otherwise.
At small volume fraction, all of the minimization procedures result in un--jammed states, and $P_N(\phi) = 0$, while conversely
at high volume fraction all of them result in jammed configurations, and $P_N(\phi) = 1$. The volume fraction dependence of 
$P_N(\phi)$ is well described by an error function
\begin{equation}
\label{eq:erf}
P_N(\phi) = \frac{1}{2}\left[1+{\rm erf} \left( \frac{\phi-\phi_N}{2\sigma_N^2} \right) \right],
\end{equation}
where $\phi_N$ and $\sigma_N$ are estimated via a least square fit. Fig.~\ref{fig:fiterf} shows the raw data for different $N$, and their fits with Eq.~\ref{eq:erf}.

\begin{figure}[t!]
\begin{center}
\includegraphics*[scale=0.30]{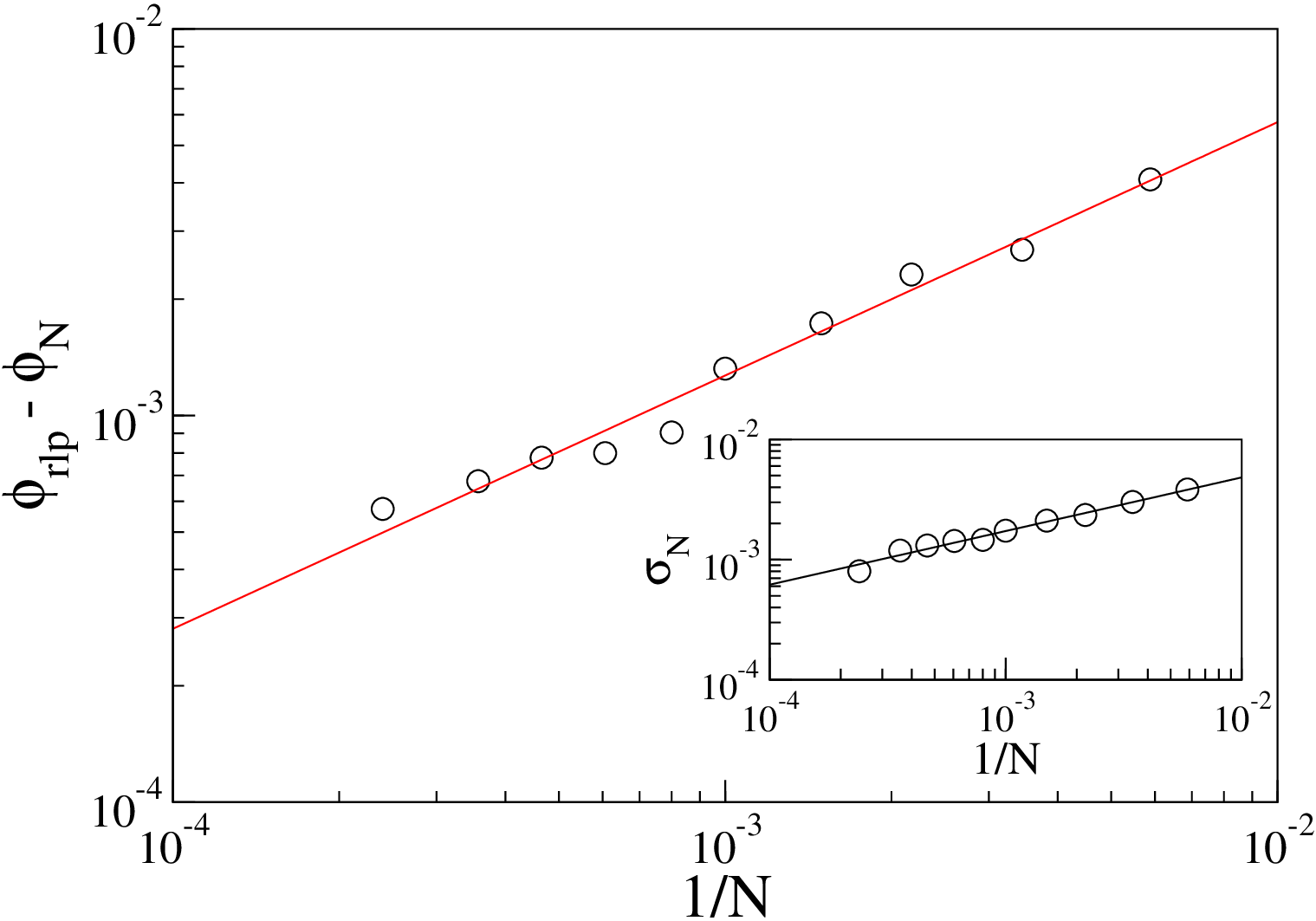}
\end{center}
\caption{\label{fig:rlp} Main panel: as the system size increases, the inflex $\phi_N$ of the $P_N(\phi)$ data approaches the asymptotic
value $\phi_\infty = \rl$ as a power law (Eq.~\ref{eq:Nrlp}). $\rl$ is well defined in the thermodynamic limit, as the width of the error
function fits $\sigma_N$ vanishes as $N$ increases (inset).}
\end{figure}

As $N$ grows, $\phi_N$ approaches an asymptotic value $\phi_\infty$, estimated\cite{OHern2003} via a power law fit of the
$\phi_N$ data. 
We show in Fig.~\ref{fig:rlp} the power law fit
\begin{equation}
\label{eq:Nrlp}
\phi_N = \phi_\infty - \Delta \phi N^{-\frac{1}{d \nu}},
\end{equation}
where $d = 3$ is the dimensionality of the system, $\phi_\infty = 0.636 \pm 0.001$ and $\nu = 0.46 \pm 0.06$. 
The value of $\phi_\infty$ is close to the one ($0.639 \pm 0.001$)  reported\cite{OHern2003} by O'Hern et al., while the discrepancy in the estimation of $\nu$ are more marked. Our estimation of $\phi_\infty$ is compatible with the smallest value of the volume fraction at which the pressure of monodisperse hard spheres has been recently found to diverge\cite{Hermes}.
$\phi_\infty$ is our best estimation of the random loose packing volume fraction in the infinite system size limit, $\rl = \phi_\infty = 0.636 \pm 0.001$. This value can be considered sharply defined, as the standard deviation $\sigma_N$ vanishes as a power law as
$N$ increases, as shown in Fig.~\ref{fig:rlp} (inset).

We note that, within our numerical accuracy an exponential law describes the data dependence of $\phi_N$ on $N$ equally well (in terms of the $\chi$-square), and provides a slightly different estimation of $\rl$, $\rl \simeq 0.6355$.

\section{Above $\rl$\label{sec:rcp}}
We have determined the value of $\rl$ finding the energy minima of the system via a procedure which is equivalent to the solution of
the equation of motion (Eq.~\ref{eq:dissipation}), in the $\eta \to \infty$ limit, using as initial state a random one.
Here we describe the results obtained when the minima of the system are obtained solving Eq.~\ref{eq:dissipation} in the presence of a
finite value of the viscosity $\eta$, until the dynamics halts.
The initial state of this minimization procedure, however, is not a true random one. This is so because the correct simulation of the
relaxation dynamics of random states, which may have very high elastic energy, requires the use of a very small numerical integration
timestep, and is therefore too computationally expensive. 
We therefore follow Zhang and Makse\cite{Makse2005}, and use the following protocol. We first prepare the system in a low volume fraction state with
zero energy (no particle contacts). The size of the particles is then quickly inflated until the desired value of the volume fraction is
reached (we increase $\phi$ linearly in time). Afterwards, Eq.~\ref{eq:dissipation} is numerically solved until the dynamics halts.
Accordingly, this procedure depends on two parameters, the rate $\Gamma$ at which the size of the particles is varied, and the viscosity
$\eta$. We use the value of $\Gamma$ considered by Zhang and Makse\cite{Makse2005}.

\begin{figure}[t!]
\begin{center}
\includegraphics*[scale=0.30]{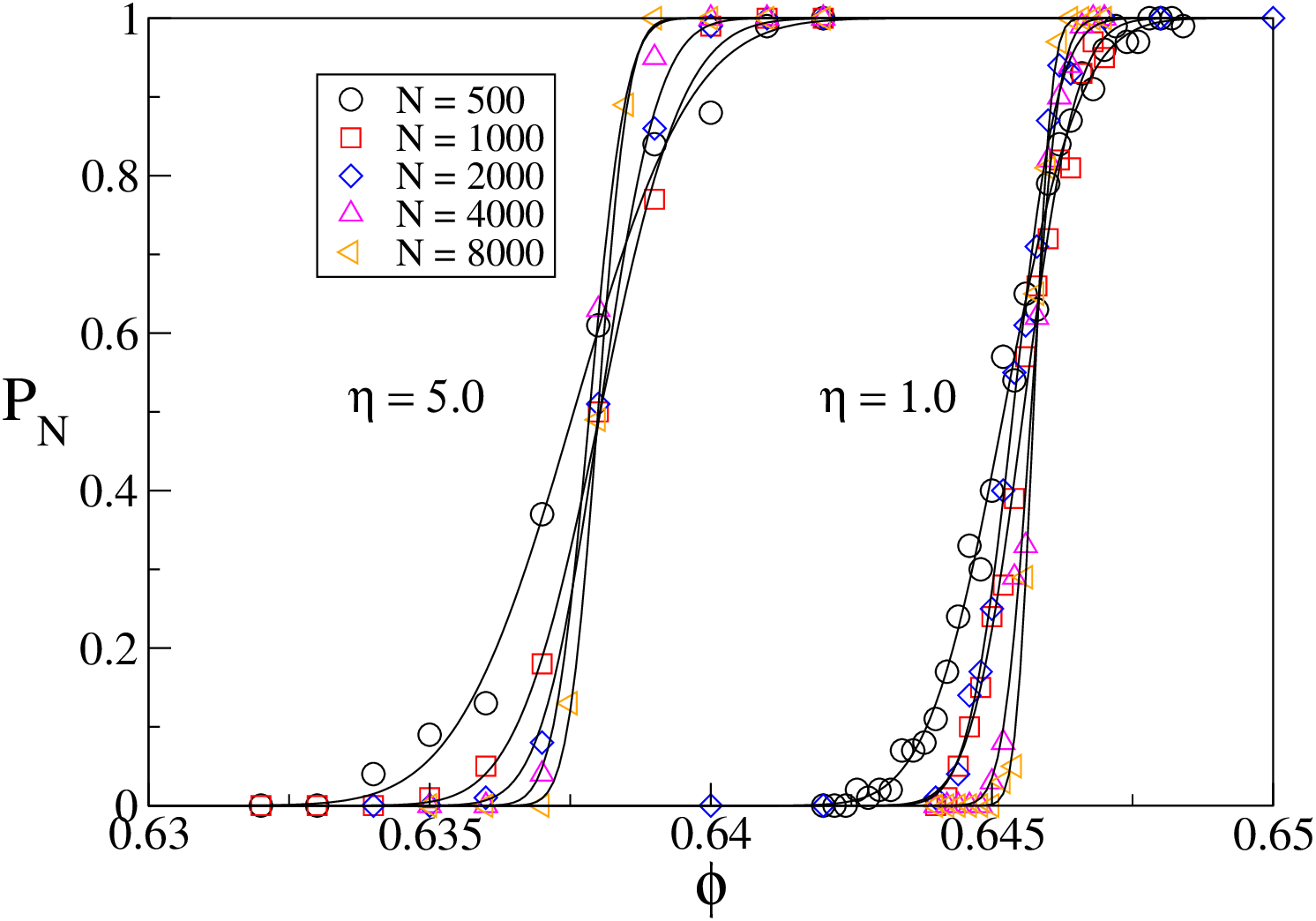}
\end{center}
\caption{\label{fig:Peta} 
The probability $P_N$ of obtaining a jammed configuration minimizing the energy of the system via the numerical solution of
Eq.~\ref{eq:dissipation}, as a function of the volume fraction $\phi$. The figure shows the results obtained for two different values of
the  viscosity $\eta$, and for several values of the system size, as indicated. Each point is estimated performing $100$ independent
minimization  procedures. Plain lines are fit to a scaled error function, Eq.~\ref{eq:erf}.}
\end{figure}

\begin{figure}[t!]
\begin{center}
\includegraphics*[scale=0.30]{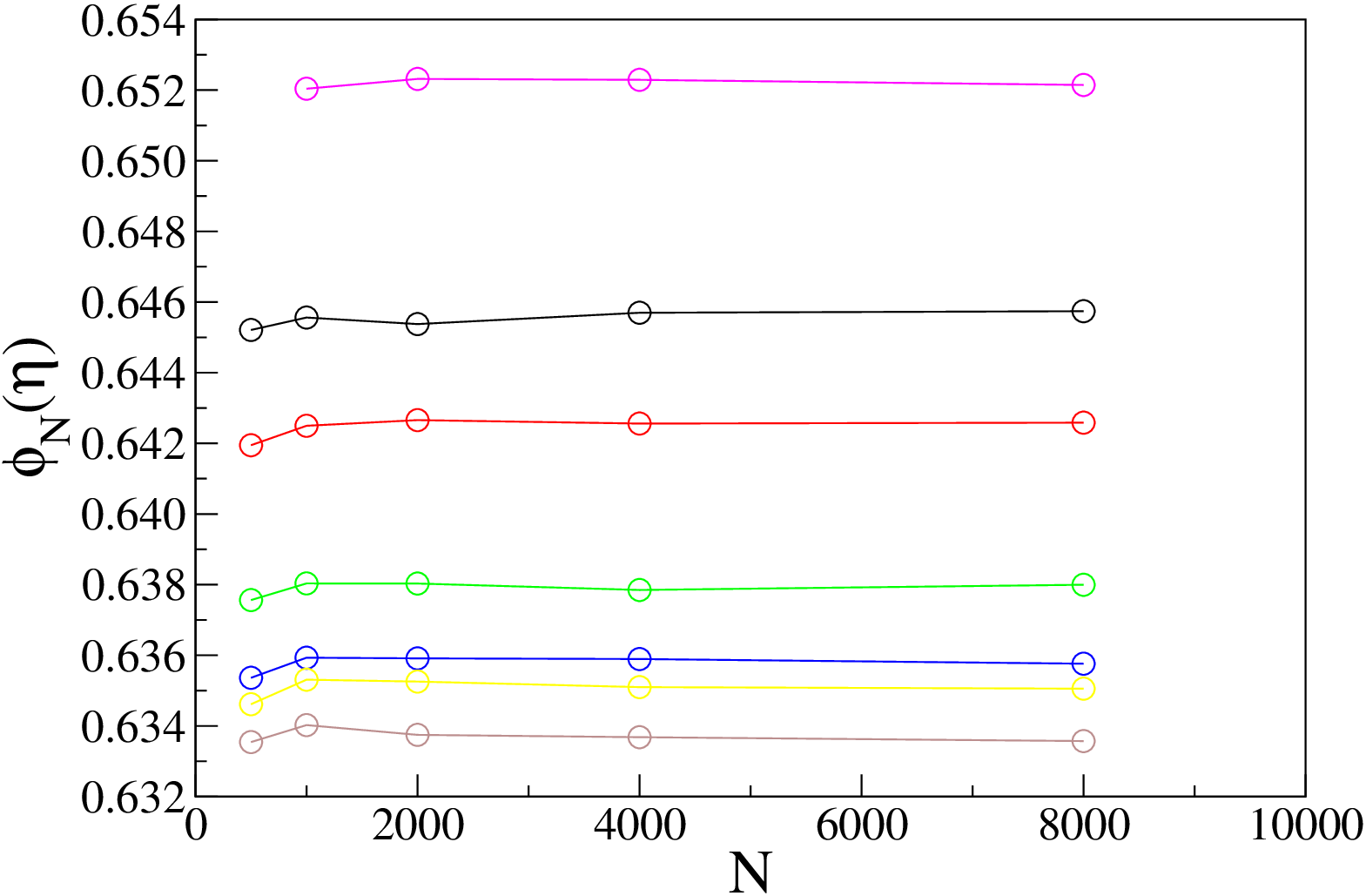}
\end{center}
\caption{\label{fig:PNeta} 
Crossover volume fraction $\phi_N$ as a function of the number of particles, for different values of the viscosity $\eta$. From bottom to
top, $\eta = 20,10,8,5,2,1,0.1$. $\phi_N$ quickly reaches an $\eta$ dependent asymptotic value as $N$ grows.}
\end{figure}

For each value of the number of particles $N$, we have considered different values of the damping parameter $\eta$, and different values of  the volume fraction $\phi$. For each $N$, $\phi$ and $\eta$ triple, we have performed $100$ simulations of the relaxation process, and determined the fraction of these simulations which resulted to be jammed, $P_N(\phi,\eta)$. We consider a system to be jammed when the elastic energy per particle is greater that $10^{-10}\varepsilon$.
Raw data for two values of $\eta$ are shown in Fig.~\ref{fig:Peta}. Fitting $P_N(\phi,\eta)$ using an error function (Eq.~\ref{eq:erf}) we
have estimated the inflection point $\phi_N(\eta)$ and the variance $\sigma_N(\eta)$. As $N$ grows, $\phi_N(\eta)$ quickly approaches its asymptotic value, as shown in Fig.~\ref{fig:PNeta}. We have therefore identified the $N \to \infty$ limit $\phi_\infty(\eta)$ with the value obtained with $N = 8000$ particles, the largest number of particles with have considered.

\begin{figure}[t!]
\begin{center}
\includegraphics*[scale=0.30]{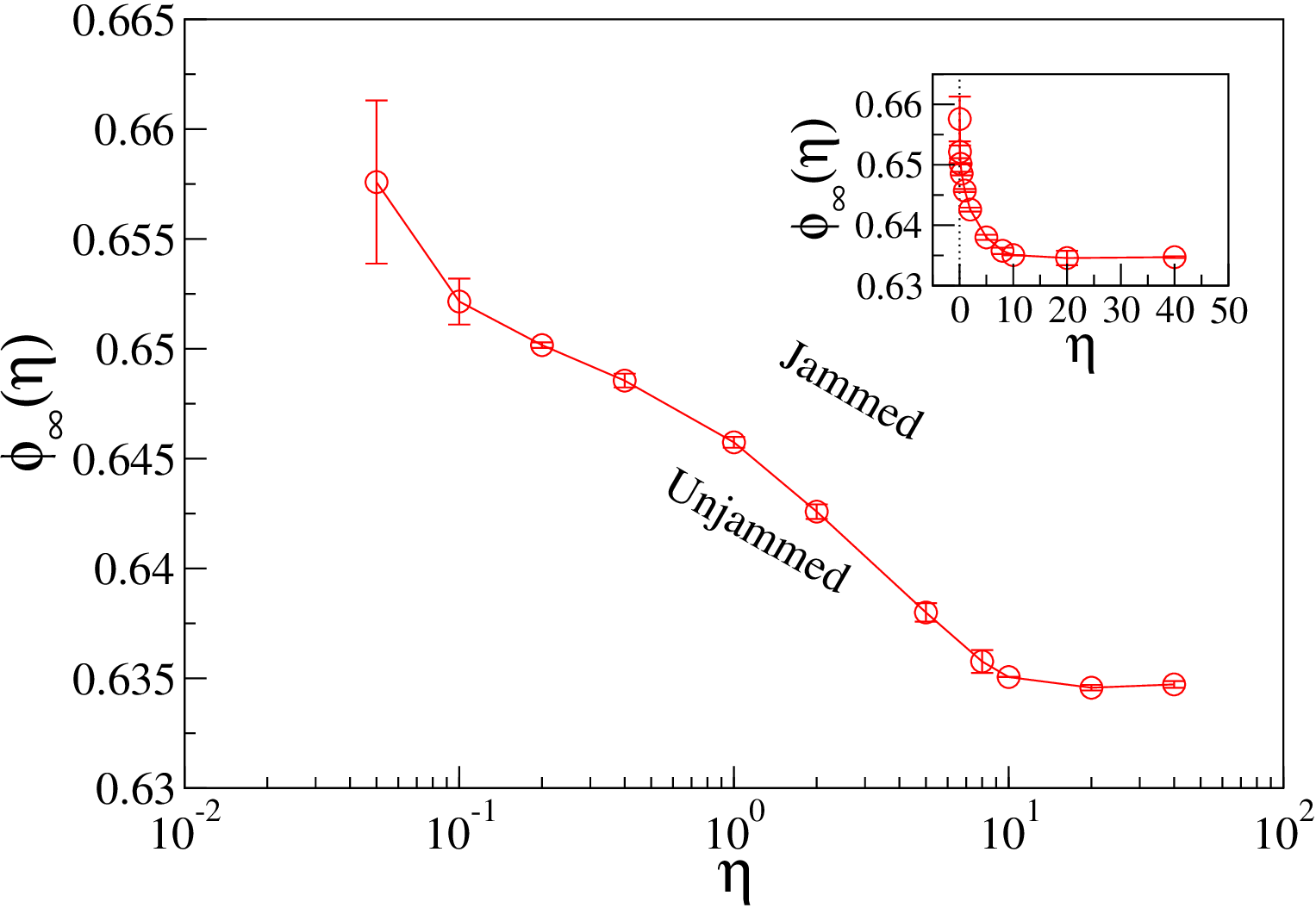}
\end{center}
\caption{\label{fig:phi_eta} 
The volume fraction $\phi_{\infty}$ of the infinite system size limit as a function of the viscosity parameter $\eta$ used in the energy
minimization procedure, in a log-linear (main panel) and in a linear scale (inset). At a given value of $\eta$, if $\phi <
\phi_{\infty}(\eta)$ the minimization procedure results in an unjammed state, while conversely a jammed state is obtained.
Here, we approximated $\phi_{\infty}$ with $\phi_{N = 8000}$. Errors are the standard deviations $\sigma_{
N = 8000}$ estimated via a least square fit of the jamming probability $P_N(\phi,\eta)$ with the error function of Eq.~\ref{eq:erf}.}
\end{figure}

The data of Fig.~\ref{fig:Peta} clearly show that by changing the viscosity parameter $\eta$ used in the minimization of the energy,
different values of the volume fraction $\phi_\infty$ are obtained. The dependence of $\phi_\infty(\eta)$
on $\eta$ is shown in Fig.~\ref{fig:phi_eta}. $\phi_\infty$ monotonously decreases as $\eta$ increases, and the limit $\eta
\to \infty$ yields $\phi_\infty(\Gamma, \eta \to \infty) = 0.634 \pm 0.001$, a value which is close to our estimation of the random loose
packing volume fraction, $\rl = 0.636 \pm 0.001$. Rigorously, one should expect $\rl = \phi_\infty(\Gamma \to \infty, \eta \to \infty)$, as
only in the $\Gamma \to \infty$ the initial state of the minimization procedure is a random one.
As $\eta$ decreases, $\phi_\infty$ increases, and reaches values which are well above any past estimate of the random close packing volume fraction. The expected emergence of ordering in these high volume fraction states is described in the next section.

\section{Random close packing\label{sec:order}}
The introduced energy minimization protocol allows the generation of unjammed packings with a volume fraction $\phi$ which is well above any past estimation of the random close packing volume fraction, as shown in Fig.~\ref{fig:phi_eta}.
Here we quantify the degree of order of these states considering systems with $N = 8000$ particles, 
which Fig.~\ref{fig:PNeta} showed to be representative of the infinite system size limit.

Following previous works\cite{Q6original, Rintoul1996,Richard1999} we quantify the degree of order focusing on the parameter $Q_l$, whose definition requires the introduction of bonds between particles. 
Here we consider two particles as bonded if they share a face of the Vorono\"i tessellation of the system. A bond between particles $i$ and $j$ defines a vector ${\bf r}_{ij} = (r,\theta,\phi)$. Each bond can be therefore associated a whole set of spherical harmonics
$Y_{l,m}(\theta,\phi)$.
By combining the values of the spherical harmonics associated to all bonds of a given particle, it is possible to associate to each particle a scalar parameter whose value depends on the shape of its Vorono\"i cell. This parameter is\cite{Q6original}
\begin{equation}
\label{eq:q6}
Q_l = \left(  \frac{4 \pi}{2l + 1}  \sum_{m = -l}^l \left|  \langle Y_{lm} \rangle \right|^2  \right)^{1/2},
\end{equation}
where the average $ \langle Y_{lm}(\theta,\phi)  \rangle$ is performed over all bonds of the particle. 
The sum over $m$ assures the independence from the chose reference frame, i.e. makes $Q_l$ rotationally invariant.
Previous works\cite{Q6original, Rintoul1996,Richard1999} have clarified that the most convenient value of $l$ 
for the study of the emergence of crystallization in systems of hard spheres is $l = 6$, which is the
lowest nonzero $Q_l$ in common with the icosahedral symmetry and cubic symmetry.
\begin{figure}[t!]
\begin{center}
\includegraphics*[scale=0.30]{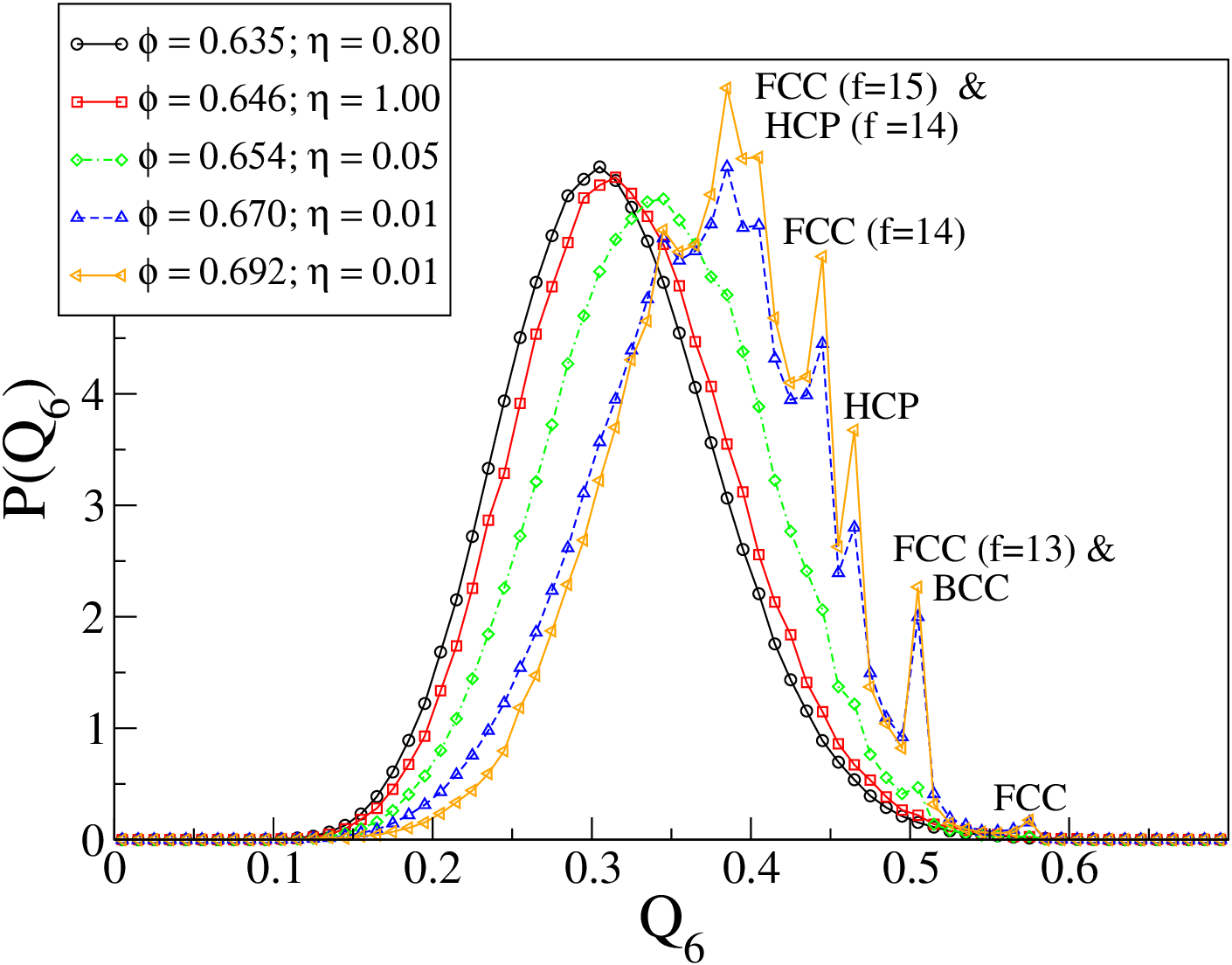}
\end{center}
\caption{\label{fig:PQ6} 
Probability distribution of the order parameter $Q_6$ of unjammed packings for $\phi = 0.635,0.646,0.654,0.670$ and $0.692$.
The curves are obtained at different values of $\eta$, as indicated. As the volume fraction increases, peaks at values of $Q_6$ known to characterize different crystalline structures appear.}
\end{figure}
Ordered structures are characterized by the frequent occurrence of typical local structures (i.e. particular shapes of the Vorono\"i cells), and are therefore characterized by a probability distribution of $P(Q_6)$ with distinct peaks. 
For instance, all Vorono\"i cells of the simple cubic crystal are equal, and $P(Q_6) = \delta(Q_6-Q_{SC})$, with $Q_{SC} \simeq 0.35$.  More complex crystals are characterized by the presence of Vorono\"i cells with few shapes, and lead to a probability distribution $P(Q_6)$ with more peaks. For this reason investigating the evolution of $P(Q_6)$ is a convenient way to monitor
the emergence of order into a system, as for instance the crystallization process of a supercooled liquid\cite{Richard1999}.
In particular, growing peaks at $Q_{HCP} \simeq 0. 48$, $Q_{BCC}\simeq 0.51$,  and $Q_{FCC} \simeq 0.57$ reveal the emergence of crystalline patches with the hexagonal close pack, the body centered cubic, and face centered cubic symmetry, respectively. It must be noted, however, that the association of a particular value of $Q_6$ to a precise local crystalline structure is not straightforward in the presence of noise. For instance, the peak at $Q_6 \simeq 0.51$ may also correspond to a FCC Vorono\"i cell with $13$ faces\cite{Richard1999}. 

We have investigated the probability $P(Q_6)$ of the unjammed configurations generated minimizing the energy with the described protocol. The distribution depends both on the volume fraction $\phi$, and on the viscous parameter $\eta$. Representative distributions are shown in Fig.~\ref{fig:PQ6}. At low volume fraction, $P(Q_6)$ has a smooth shape, while peaks are clearly present at high volume fraction. These peaks signal the occurrence of local arrangements typical of the HCP, of the BCC (or FCC with 13 faces) and of the FCC crystal. Other peaks correspond to the presence of many nearly-cristalline Vorono\"i cells.

\begin{figure}[t!]
\begin{center}
\includegraphics*[scale=0.30]{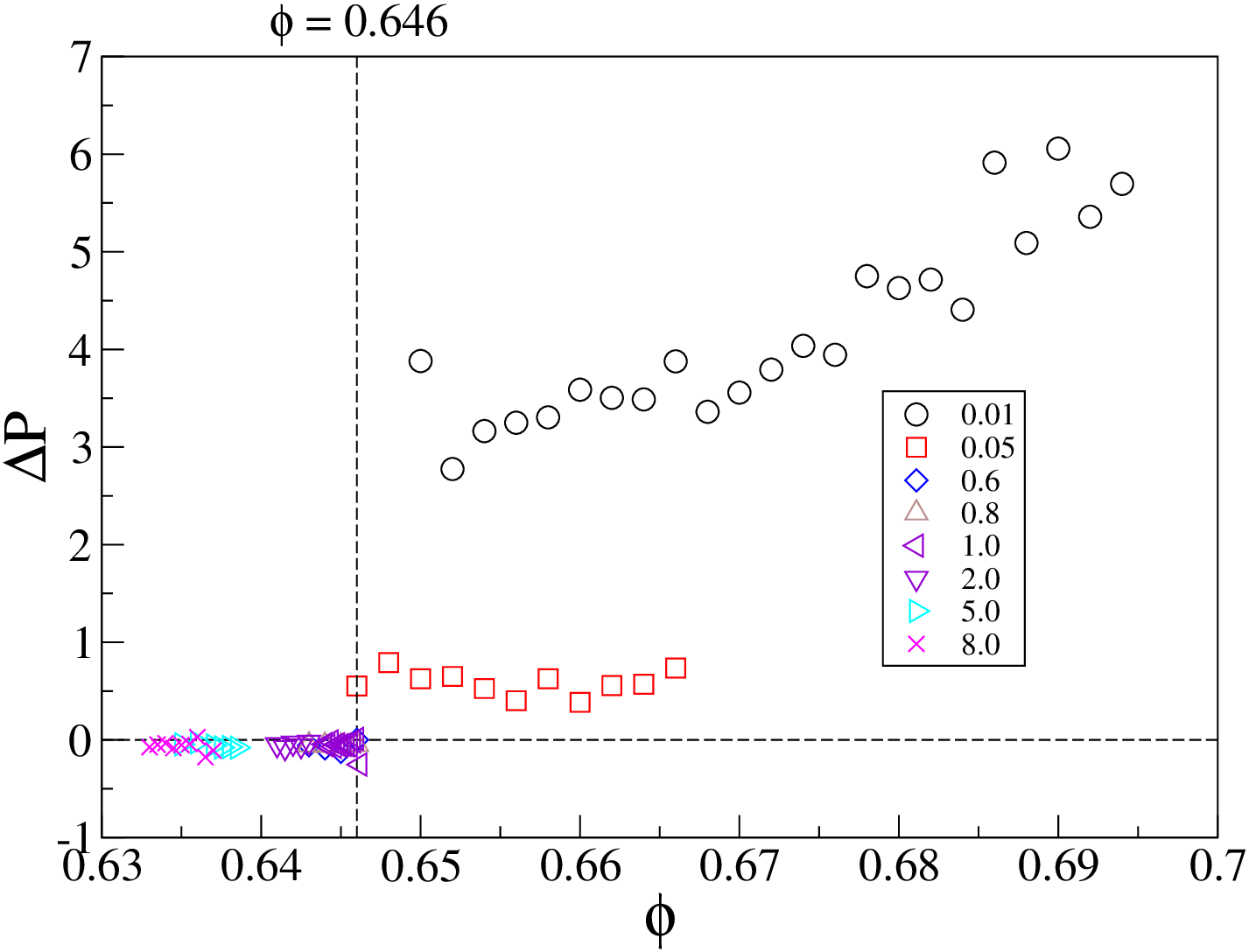}
\end{center}
\caption{\label{fig:DQ6} 
Order parameter $\Delta P$ of unjammed states as a function of the volume fraction, for different values of the viscous parameter  $\eta = 0.01$, $0.05$, $0.6$, $0.8$, $1.0$, $5.0$ and $8.0$ used in the minimization procedure. For $\phi < 0.646$, there are values of $\eta$ leading to disordered unjammed states with $\Delta P \simeq 0$. Conversely, for $\phi > 0.646$ we find $\Delta P > 0$ for all values of $\eta$, suggesting that all unjammed states with $\phi > 0.646$ have crystalline patches.}
\end{figure}

From the probability distribution $P(Q_6)$ it is possible to extract a scalar order parameter which quantifies the degree of order by measuring the hight of the peaks with respect to the `base' of the distribution. Here we suggest to first identify the $n_p$ values $Q^p_k$, $k = 1,\ldots n_p$ where the distribution $P(Q_6)$ has peaks, and then to define the order parameter as 
\begin{equation}
 \Delta P = \sum_{k=1}^{n_p} P(Q^p_k) - \frac{1}{2} \left[ P(Q_{k}^p+ \delta Q) + P(Q_{k}^p - \delta Q) \right]
\end{equation}
where the value of $\delta Q$ is irrelevant as long as this is larger than the width of the peaks, and smaller than the distance between consecutive peaks. 
This order parameter compares the height of the peaks of the $P(Q_6)$ distribution at values of $Q_6$ characterizing the ordered structures with estimates obtained via a local linear approaximation of the distribution. 
If the distribution has no peaks signaling the presence of ordered structures, then $\Delta P \simeq 0$, while $\Delta P > 0$ signal the presence of crystalline patches. 
We have computed $\Delta P$ fixing $\delta Q = 0.1$, and $n_p = 3$, where $Q^p_1 = Q_{hcp}$,  $Q^p_2 = Q_{bcc}$, $Q^p_3 = Q_{fcc}$. The inclusiong of peaks observed at smaller $Q_6$ (e.g $Q_6 \simeq 0.43$ or $Q_6 \simeq 0.38$) is irrelevant, as these peaks correspond to slightly irregular crystalline Vorono\"i cells, and are never observed alone.

The dependence of $\Delta P$ on the volume fraction, for different values of $\eta$, is shown Fig.~\ref{fig:DQ6}, and reveals the presence of a transition occurring at a volume fraction $\phi \simeq 0.646$. Regardless of the value of $\eta$, all unjammed states with $\phi \gtrsim 0.646$ have some degree of order. Conversely, when $\phi \lesssim 0.646$, high values of $\eta$ lead to disordered unjammed states.

\begin{figure}[t!]
\begin{center}
\includegraphics*[scale=0.30]{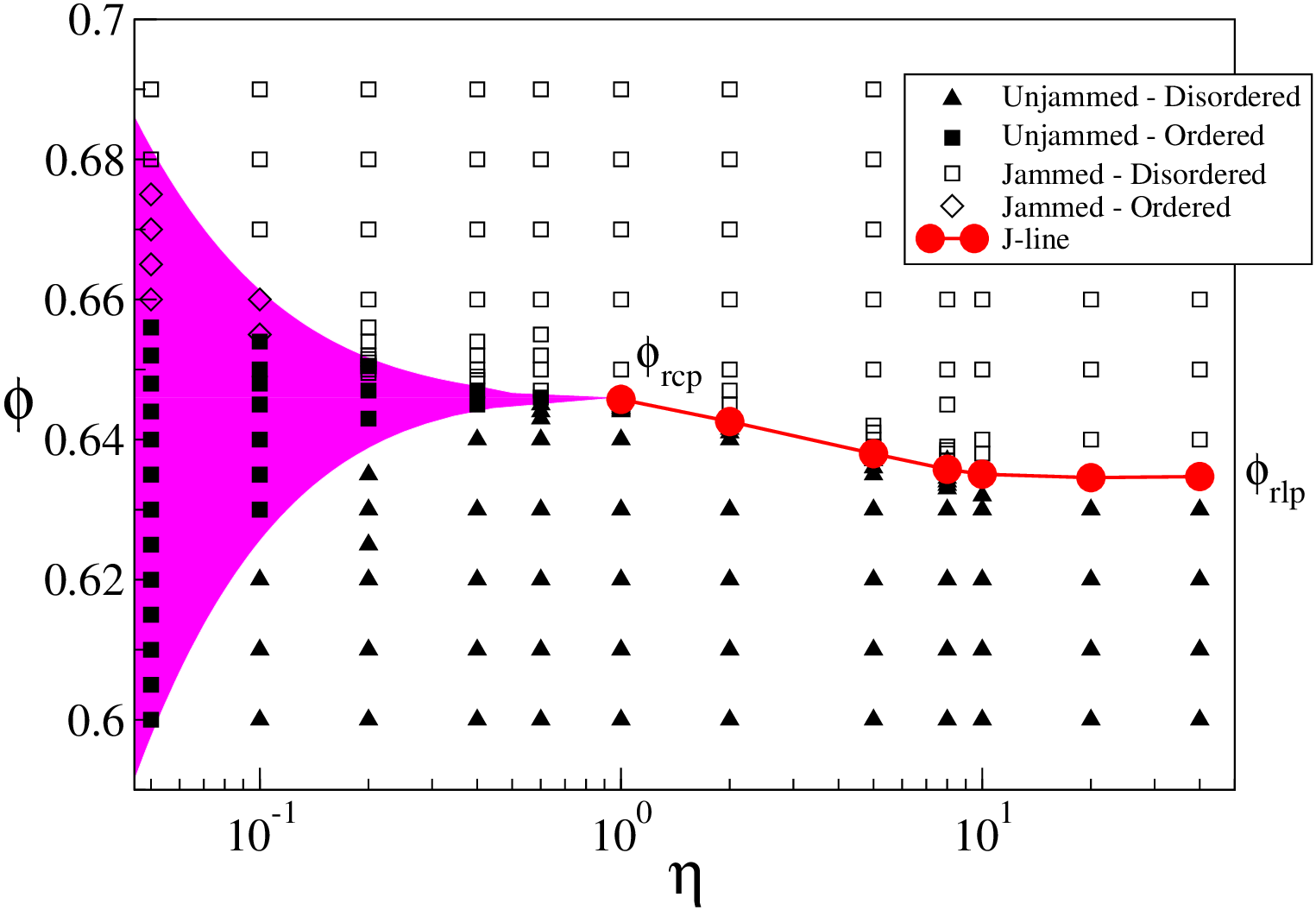}
\end{center}
\caption{\label{fig:rc_rl} 
Jamming and ordering properties as a function of the volume fraction and of the viscosity parameter.
The shaded area covers the region of the $\phi$--$\eta$ plane where crystalline patches 
are found in the generated packings. The large full circles identify the jamming transition line 
between fluid like disordered states and disordered solid states. 
The extrema of this line are the random loose ($\eta \to \infty$) and
the random close packing volume fraction, respectively.
}
\end{figure}

The jamming and the ordering properties of the system are summarizied in the $\phi$--$\eta$ diagram shown in
Fig.~\ref{fig:rc_rl}. This diagram clarifies that the jamming transition bewteen disordered states occurs in a volume fraction range,
with extrema $\rl \simeq 0.635$ and $\rc \simeq 0.646$. The jamming transition line also extends to values of $\phi > \rc$, as shown in Fig.~\ref{fig:phi_eta}, but at this high valued of the volume fraction ordering is found in the separated unjammed and/or jammed states. Fig.~\ref{fig:rc_rl} clarifies that the tendency towards the formation of crystalline patches is higher the smaller the viscosity parameter. In this sense, the conjugate gradient protocol\cite{OHern2002,OHern2003} is the lesser prone towards crystallization. The question\cite{rvlp} whereas there are jammed states below $\rl$, which may be obtained using different protocols, will be discussed in the next session.

\section{Discussion\label{sec:conclusions}}
We have introduced a protocol able to generate jammed zero pressure disordered packings of frictionless spheres with a volume fraction varying in a whole range. The obtained volume fraction depends on the value of a viscosity parameter $\eta$.
The lower extremum of this volume fraction range is obtained solving the equations of motion of the system in the quasistatic limit $\eta \to \infty$, starting from a random initial condition. Following Onoda and Liniger\cite{Onoda1990}, the quasistatic minimization of the energy can be considered as the operative definition of the random loose volume fraction. Our results suggest $\rl \simeq 0.636$.

As the viscosity parameter $\eta$ decreases, the volume fraction of the jammed configurations increases. 
Introducing an order parameter based on the probability distribution of finding Vorono\"i cells with peculiar shapes, we show that our
numerical protocols generate un--jammed disordered states up to a volume fraction $\rc \simeq 0.645$, we identified with the random close packing volume fraction. We cannot exclude that its value may change if one considers a different order parameter.
For $\phi > \rc$ all unjammed states appear to contain crystalline patches. Zero pressure disordered jammed
frictionless spheres can therefore be found  in a whole volume fraction range, at least extending from $\rl$ to $\rc$. 
This scenario is consistent with recent numerical results investigating the jamming transition of thermal systems~\cite{Hermes,Sastry}.

\begin{figure}[t!]
\begin{center}
\includegraphics*[scale=0.30]{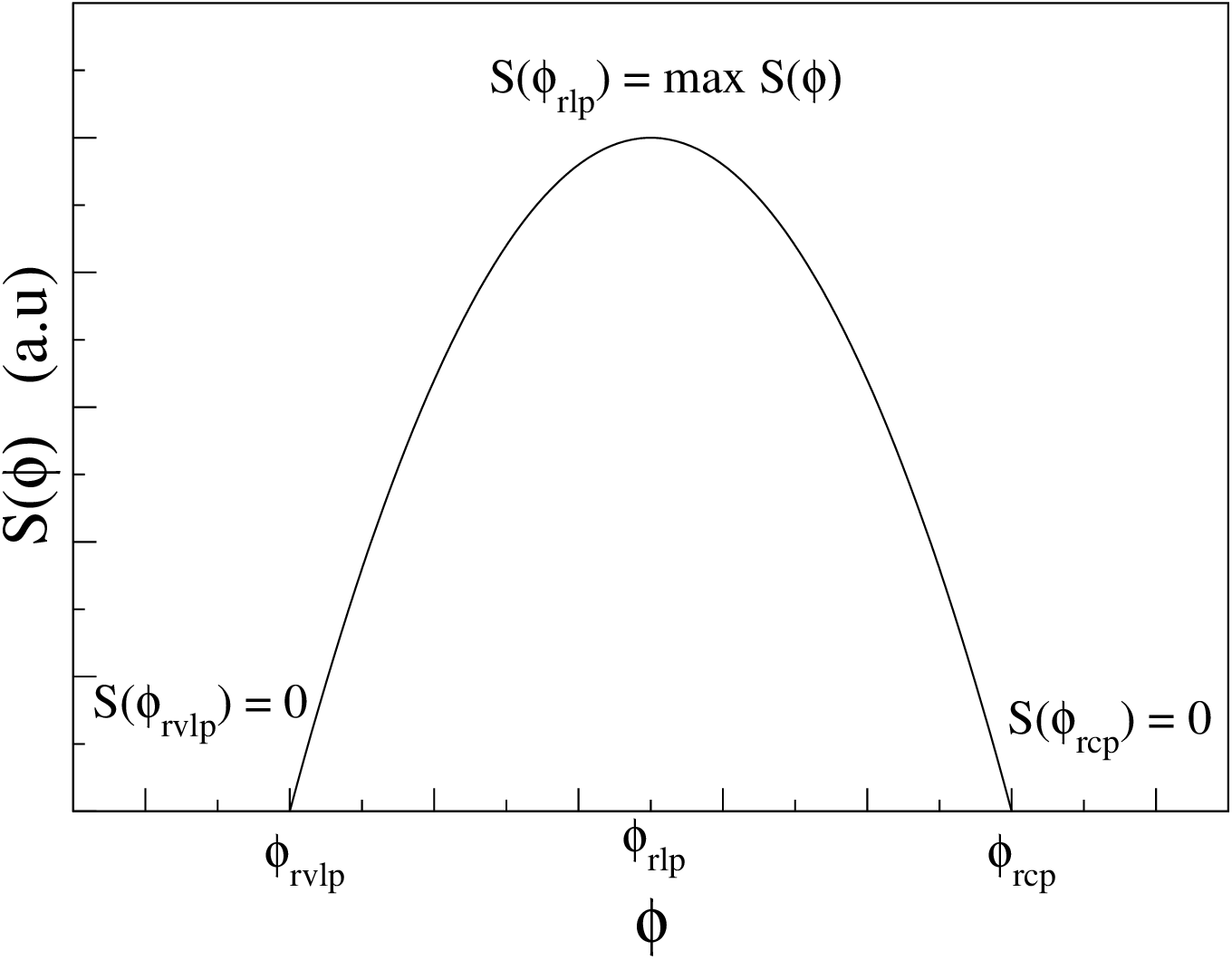}
\end{center}
\caption{\label{fig:entropy} 
Schematic~\cite{rvlp} dependence of the entropy $S(\phi) = \log \Omega(\phi)$ on the volume fraction, where $\Omega(\phi)$ is the density of disordered jammed states.
}
\end{figure}

An insight on the origin of these limiting values of the volume fraction is obtained considering how many different jammed zero pressure states exists at any given value of the volume fraction $\phi$.
This leads to the introduction of the density of disordered jammed zero-pressure states\cite{rvlp,Makse2009}, $\Omega(\phi)$, or of the entropy $S(\phi) = \log \Omega(\phi)$.  We have recently investigated this quantity in a two dimensional model\cite{rvlp}, finding an non-monotonous entropy $S(\phi)$ which is zero at small $\phi$, then increases with $\phi$ up to its maximum value, and finally decreases with $\phi$ until it vanishes at high $\phi$, as scheamtically illustrated in Fig.~\ref{fig:entropy}.

When minimizing the energy in the infinite vicosity limit starting from random initial configurations, it is reasonable to assume as a first approximation that one finds all jammed states with the same probability (more precisely, one should consider that the probability of finding a minima is proportional to the width of its energy basin). If this is so, then the volume fraction obtained in the $\eta \to \infty$ limit is the one where the entropy $S(\phi)$ has a maximum, which should be therefore identified with the random loose packing volume fraction. Likewise, since our results suggest that there are not disordered unjammed packing with $\phi > \rc$, then $\rc$ could be associated with the value of the volume fraction where the entropy vanishes, at high volume fraction. 
Accordingly, the non-monotonous variation of the entropy with the volume fraction\cite{rvlp} indicates that there exist disordered zero pressure jammed states with volume fraction $\phi < \rl$. The smallest volume fraction of these states should correspond to the random very loose volume fraction\cite{rvlp}, $\vrl$.

Open questions ahead include the developing of protocols to generate disordered jammed states of with volume fraction $\phi < \rl$,  as well as the clarification of the volume fraction dependence of mechanical and geometrical properties of jammed packings.

\bibliographystyle{rsc}


\end{document}